# Performance of ChatGPT on tasks involving physics visual representations: the case of the Brief Electricity and Magnetism Assessment

Giulia Polverini, Jakob Melin, Elias Önerud, and Bor Gregorcic

Department of Physics and Astronomy, Uppsala University, Box 516, 75120 Uppsala, Sweden

**Abstract:** Artificial intelligence-based chatbots are increasingly influencing physics education due to their ability to interpret and respond to textual and visual inputs. This study evaluates the performance of two large multimodal model-based chatbots, ChatGPT-4 and ChatGPT-4o on the Brief Electricity and Magnetism Assessment (BEMA), a conceptual physics inventory rich in visual representations such as vector fields, circuit diagrams, and graphs. Quantitative analysis shows that ChatGPT-4o outperforms both ChatGPT-4 and a large sample of university students, and demonstrates improvements in ChatGPT-4o's vision interpretation ability over its predecessor ChatGPT-4. However, qualitative analysis of ChatGPT-4o's responses reveals persistent challenges. We identified three types of difficulties in the chatbot's responses to tasks on BEMA: (1) difficulties with visual interpretation, (2) difficulties in providing correct physics laws or rules, and (3) difficulties with spatial coordination and application of physics representations. Spatial reasoning tasks, particularly those requiring the use of the right-hand rule, proved especially problematic. These findings highlight that the most broadly used large multimodal model-based chatbot, ChatGPT-4o, still exhibits significant difficulties in engaging with physics tasks involving visual representations. While the chatbot shows potential for educational applications, including personalized tutoring and accessibility support for students who are blind or have low vision, its limitations necessitate caution. On the other hand, our findings can also be leveraged to design assessments that are difficult for chatbots to solve.

## I. INTRODUCTION

Large Language Models (LLMs) have emerged as a transformative force in society since late 2022, when OpenAI's GPT model was made available to the public in the form of ChatGPT, a user-friendly and broadly accessible chatbot. LLM-based chatbots can process textual inputs, including natural language, mathematical symbols, and computer code, and generate contextually relevant textual outputs. Their abilities, in combination with ease of use and broad availability, have disrupted various areas of life, with education being one of the most impacted [1,2]. The rapid adoption, quickly improving capabilities, and inherent flexibility of their use make LLMs a technology that continues to influence and disrupt educational practices worldwide [3].

Physics education is no exception to this trend [4]. A significant advancement in LLM technology has been the development of models able to process not only text but also other data types, such as images and audio. These are commonly referred to as Large Multimodal Models (LMMs) or Multimodal Large Language Models (MLLMs). Their ability to interpret diverse inputs broadens their potential for educational applications. In the context of physics education, the relevance of LMMs is particularly pronounced, because doing physics requires a deep engagement with diagrams, graphs, sketches [5], as well as other visual and embodied representations [6,7].

Because building LLMs and LMMs from scratch and fine-tuning them is computationally demanding, costly, and requires large amounts of data, educators often turn to commercial actors like OpenAI and Google for ready-made baseline models [8]. These models serve as foundational tools that can be adapted for a variety of educational purposes (e.g., [9–11]). Given their growing adoption and diverse potential applications, it is important for the educational research community to continuously assess the abilities of widely used commercial LLMs and LMMs in different educational contexts. The ability to work with different graphical representations, which is especially relevant for physics education, is one such area in need of continuous assessment.

Our study contributes to this relatively young research field by investigating the ability of widely used LMM-based chatbots – ChatGPT-4 and ChatGPT-4o – to engage with conceptual physics tasks involving diverse graphical representations. This is a continuation of our previous work assessing ChatGPT's ability to interpret kinematics graphs [12], but extends our scope to a broader range of physics representations that have not been systematically explored yet. To do this, we focus on electricity and magnetism, an area of physics particularly rich in different representations. We test OpenAI chatbots' performance on the Brief Electricity and Magnetism Assessment (BEMA) [13], a well-established research-based survey in physics education research, which features a diverse range of disciplinary representations, including sketches, vectors, vector fields, circuit diagrams, and graphs.

The study addresses the following two research questions:

*RQ1. What is the performance of ChatGPT-4 and ChatGPT-4o on BEMA?*

*RQ2. What difficulties does the more recent and more widely used model, ChatGPT-4o, face in interpreting and solving tasks that involve diverse visual physics representations?*

To answer these questions, we divide our investigation into two parts. For RQ1, we quantitatively evaluate the performance of both models on BEMA, comparing their performance to each other and to those of a sample of university students taking a calculus-based course on electricity and magnetism [14]. For RQ2, we investigate qualitatively the difficulties encountered by the more recent and more widely used model, ChatGPT-4o, in solving tasks on BEMA. By addressing these research questions, we aim to provide the physics education community with a critical perspective on the abilities of the two broadly used commercial LLM-based chatbots. Our findings can guide decisions on whether and how these models can effectively support the teaching and learning of representationally rich physics topics. For instance, our findings have the potential to inform the design of educational applications and accessibility solutions for learners who are blind or have low vision, as well as support efforts to uphold academic integrity through the design and selection of assessments, which is difficult for chatbots to answer correctly. Furthermore, our analysis of the difficulties exhibited by the studied LMMs provides researchers with a possible approach for investigating LMMs abilities in the future. Because the technology is advancing at a very rapid pace, we expect that such evaluations will be necessary at relatively short time intervals.

## II. BACKGROUND

The integration of LLMs and LMMs into physics education has sparked a growing body of research aimed at evaluating their capabilities and exploring their potential applications. As these AI models continue to evolve, researchers are investigating both their performance on physics-related tasks and the ways in which they can be integrated into the educational process to support teaching and learning. This section provides an overview of these two key aspects.

### A. Performance of LLMs and LMMs in physics education

Since the public release of ChatGPT in November 2022, and the implementation of its vision ability in October 2023, LLMs and LMMs have become active research areas in physics education. Early studies have laid the groundwork for understanding how these models perform on various physics tasks, uncovering both opportunities and challenges in leveraging them for educational purposes. Research has evaluated their performance across a range of physics contexts:

*Research-based assessments.* Several studies have assessed LLMs' performance on the Force Concept Inventory (FCI). ChatGPT-3.5 was shown to perform at beginners' level [15,16], while ChatGPT-4 demonstrated expert-level performance [17]. In addition, Wheeler and Scherr [18] found that ChatGPT-3.5's responses to FCI often mirrored common student misconceptions. Kieser et al. [19] found that ChatGPT-4 achieved 83% correctness on the test and was, with appropriate prompting, also able to simulate realistic student misconceptions, offering a potential tool for the refining of surveys such as FCI. While early studies provided the chatbot with textual descriptions of the images on the test [15,16], Aldazharova *et al*. [20] explored how ChatGPT performs when it is given the images as part of the multimodal input. While the model demonstrated strong reasoning abilities, often providing logical explanations even for incorrect answers, it struggled significantly with spatial reasoning and figure-based questions. Kortemeyer et al. [21] analyzed GPT-4o's performance across multiple physics concept inventories across different languages. Their findings highlight significant variation in performance, with the AI excelling in English, Spanish, and Portuguese but struggling in Punjabi and Tamil. Additionally, they found that the AI consistently underperforms on questions requiring interpretation of images and laboratory-based reasoning.

*Conceptual physics tasks.* Gregorcic and Pendrill [22] analyzed ChatGPT-3.5's responses to a simple conceptual mechanics problem and found that its answers often contained incorrect or contradictory reasoning. In a follow-up study, Gregorcic et al. [23] show that ChatGPT-4 has significantly improved in engaging in Socratic dialogue, making it more responsive to guided questioning and better at revising its own reasoning when inconsistencies are pointed out. However, in a previous version of this study [24], the authors highlighted a persistent weakness in AI-generated graphical representations. When prompted to generate a velocity-time graph for a bouncing ball, ChatGPT-4 attempted to write and execute Python code but produced incorrect outputs, misrepresenting graph features such as slopes and discontinuities. Despite these improvements, Polverini and Gregorcic [25] found that ChatGPT-4's performance on conceptual physics tasks remains far from perfect, often displaying reasoning inconsistencies and errors. However, they demonstrated that prompt engineering techniques, such as domain specification and Chain-of-Thought reasoning (see section II.C), can enhance the quality of the chatbot's responses. Sirnoorkar et al. [26] explored the responses of ChatGPT-3.5 and 4 to physics problems framed to elicit sensemaking and mechanistic reasoning. While the LLMs' responses showcased structured and detailed solutions, students demonstrated comparatively richer epistemic practices such as iterative refinement and diagram-based reasoning.

*Short-form physics essay-writing.* Yeadon et al. [27] demonstrated that ChatGPT-3.5 could produce essays in a university course on the history and philosophy of physics that would achieve first-class grades. Additionally, in a follow-up study [28], the authors suggest that at-home essay-style examination is no longer sensible, since it is very hard to reliably detect LLM-generated text when it is even slightly modified by the user. Similarly, Bernabei et al. [29] found that AI-assisted essays were of comparable quality to non-AI-assisted ones and that existing AI detection tools were largely ineffective.

*Programming tasks.* Kortemeyer [15] found that ChatGPT-3.5 is very good at solving programming-related tasks, which often require a logical and structured approach. Yeadon et al. [30] explored the performance of GPT-3.5 and 4 in university-level physics coding assignments, revealing that while GPT-4 can approach good performance, student submissions were often better due to nuanced and creative design choices.

*Laboratory-based tasks.* Low and Kalender [31] tested ChatGPT-4 in combination with a Python code-running plugin in introductory mechanics lab-related tasks. They assessed its ability to generate realistic synthetic data, analyze real experimental data, and perform statistical calculations. They found that while ChatGPT-4 could successfully handle such tasks, the quality of its output heavily depended on the specificity of prompts, and its synthetic data often displayed heteroscedasticity. Kilde-Westberg et al. [32] examined how high school students used ChatGPT-3.5 as a lab partner in a physics experiment on acoustic levitation and the speed of sound. Their findings highlight that the chatbot can aid conceptual understanding and provide relevant formulas but is unreliable for calculations and requires critical oversight from students and teachers.

*Physics course examinations.* Yeadon and Hardy [33] evaluated ChatGPT-3.5 on physics exam questions across secondary school and university levels. While it mostly provided correct answers for simpler problems, its performance varied significantly with problem complexity. Similarly, Frenkel and Emara [34] showed that GPT-4 achieved a correctness rate of 76.6% on engineering exams, significantly outperforming GPT-3.5's 51.1%, but struggled with multistep calculations. More recently, Pimbblet and Morrell [35] investigated ChatGPT-4's performance on summative examinations for all courses leading to an undergraduate physics degree at a UK-based university. The chatbot achieved an upper second-class grade overall (65%), by excelling in computational tasks and single-step problems. However, it had difficulties handling multi-step reasoning, graphical interpretation, and laboratory or viva-based assessments.

*Problem-solving.* Wang et al. [36] explored ChatGPT-4's ability to solve problems from a college-level engineering physics course, showing that the chatbot can achieve a success rate of 62.5% on well-specified problems but struggled with under-specified problems. Kumar and Kats [37] demonstrated that ChatGPT-4, in combination with a Python code-running plugin, could successfully solve introductory college-level vector calculus and electromagnetism problems.

*Image interpretation.* Suhonen [38] evaluated multiple LLM-based chatbots in solving thermodynamics and mechanics problems. While ChatGPT-4 was the most capable, and successfully generated a temperature vs. time graph and correctly identified key physics principles, it still made certain content errors related to slopes and algebraic expressions. Additionally, the chatbots struggled with creating accurate free-body diagrams. Polverini and Gregorcic [12] tested ChatGPT-4's ability to interpret kinematics graphs and found that while its textual reasoning was often correct, its visual interpretation of graphs was flawed. When textual transcriptions replaced graphical inputs, performance improved significantly, suggesting limitations in the chatbot's visual processing abilities. In follow-up studies [39,40], the authors evaluated a broader range of chatbots, both freely available and subscription-based. They attempted to categorize tasks to determine which were easier or harder for the LMMs, but no meaningful pattern or category emerged. The only clear distinction was that linguistic tasks were consistently easier for chatbots than tasks requiring visual interpretation. Enabling ChatGPT-4 to use the Python code-running plugin[1] did not result in an improvement in the overall performance on the studied image interpretation tasks.

Collectively, these studies indicate that LLMs and LMMs mostly excel in well-defined, structured problems that involve natural language processing, namely straightforward applications of formulas or principles, regardless of the specific subarea of physics, as well as in physics programming tasks. However, they often struggle more with multi-step reasoning or problems requiring implicit assumptions or contextual understanding.

In this regard, recent developments of specialized reasoning models, such as OpenAI's o1 [41] and o3 [42], as well as DeepSeek's R1 [43] suggest that tasks involving complex reasoning may soon become more manageable for these systems [44]. These models use specialized approaches to answering questions, often referred to as Chain-of-Thought (CoT) (see section II.C).
Studies also show that the chatbots' multimodal ability to process and generate images is comparably less advanced than their linguistic capabilities. However, the limited research on LMMs' visual interpretation abilities leaves much to be explored in this area. This paper aims to contribute to this emerging field by assessing their effectiveness in handling visual input, a crucial factor in their potential adoption in physics education. Given physics' heavy reliance on graphical representations [45], understanding how well LMMs interpret and integrate visual information into problem-solving can be of great interest to the physics education community.

[1] Advanced Data Analysis, previously known as Code Interpreter.

### B. A framework for potential uses of LLMs and LMMs in physics education

The growing interest in integrating LLMs and LMMs into education has led to several proposals aimed at enhancing the teaching and learning of physics. Against the background of existing literature on educational conceptualizations and applications of this technology in physics, we propose to use Taylor's framework [46] as a guide for thinking about possible AI uses. In his seminal work on the use of computers in education, Taylor conceptualized three roles that computers could play in the education process: tutors, tutees, and tools. Inspired by his work, we appropriate his framework and use it to describe some possible applications of LLMs and LMMs in the context of physics education.

#### *1. Tutor*

The concept of AI serving as teachers dates back several decades, with speculations emerging as early as the 1950s. Alan Turing imagined machines that could eventually learn and simulate human-like intelligence, implicitly raising the question of whether they could assume the role of educators [47]. Since then, research has been conducted on AI-driven education, from early intelligent tutoring systems [48,49] to adaptive learning algorithms [50], culminating in the present era where LLM- and LMM-based chatbots are actively being tested as interactive tutors, capable of engaging in real-time dialogue and personalized instruction [51].
Researchers have been developing institutional tutoring systems [9–11] as well as exploring the potential of chatbots to take on different tutoring roles, such as tips [52] and feedback [53,54] providers. The evolution of LLMs into LMMs has also led large AI companies to promote LMM-based chatbots as advanced physics tutoring systems [55,56].

#### *2. Model of a student*

In Taylor's original formulation, the tutee role referred to the idea of teaching the computer, namely programming. In the context of AI, we reinterpret this role to mean AI as "a model of a student," in the sense that it can simulate student behavior rather than merely being taught in the traditional sense. LLMs and LMMs are being used in physics education research to mimic student misconceptions, reasoning, and problem-solving approaches [18,19]. In addition, this adaptation of the tutee role offers applications in teacher training, allowing educators to practice Socratic dialogue, anticipate common difficulties, and refine their instructional strategies [22,23,57].

#### *3. Tool*

AI can act as an assistive technology to support both teaching and learning. The tool role positions LLM- and LMM-based chatbots as a resource that students and teachers can leverage to improve educational experiences and outcomes.
For teachers, chatbots hold the potential to transform classrooms by streamlining tasks that would otherwise consume valuable time. A chatbot can become an assistant for grading students' submissions, both handwritten [58,59] and digital [60], as well as designing lesson plans and developing problem sets [57,61]. Notably, assisting teachers with grading and feedback has garnered significant interest, particularly given the increasing demands on educators, who often face challenges such as large class sizes, limited resources, and high costs associated with education [62].

For students, chatbots can serve as interactive peers to "think with" [63], a partner for lab activities [32,64,65], or coding [66], and develop deeper conceptual understanding [67]. The textual and image output they generate can serve as material for practicing critical reading and thinking [68,69]. LMMs also hold promise as accessibility tools, potentially offering students with low vision assistance with interpreting visual information [65,70].

### C. Chain of Thought

Our previous research [25] and the data collected for the current study [71] show that ChatGPT-4 and 4o nearly always provide argumentation to justify their final answers. This feature results from the use of the Chain-of-Thought approach (CoT), a technique that directs a chatbot to write out its reasoning step-by-step [72]. By writing out the answer step-by-step, CoT enables the chatbot to present what appears to be a structured argument supporting its conclusion. This technique has proven effective in improving LLMs' performance on tasks involving multiple and complex reasoning steps [73]. Initially, explicit CoT prompting was necessary to elicit such reasoning [25]. However, recent developments suggest that OpenAI has integrated CoT into ChatGPT-4's and 4o's default pre-prompt, making step-by-step reasoning a standard behavior without requiring special CoT prompting from the user. Specialized models, such as OpenAI's o1 [41] and o3 [42], or DeepSeek's R1 [43], were designed and trained to leverage more advanced forms of CoT for solving complex reasoning problems, showing significant advancements in reasoning benchmark evaluations (e.g., [44]).

This built-in reasoning process not only enhances the chatbot's performance on complex tasks but also serves as a valuable tool for researchers. Examining the generated reasoning steps provides a window into the models' problem-solving process, allowing researchers to qualitatively assess their capabilities and potential limitations (see [74], for example). However, most existing studies have focused on textual or numerical problems, with little exploration of CoT in tasks requiring the integration of textual and graphical reasoning.

This paper addresses this gap by investigating LMM-based chatbots' performance on physics tasks that combine textual and visual reasoning. It also contributes to the emerging literature on CoT analysis in the context of multimodal physics tasks, with a focus on tasks involving visual representations of electricity and magnetism. In section V, we take advantage of the CoT style of responses to infer ChatGPT-4o's abilities by closely analyzing its output. Since technical details about how ChatGPT processes images are not disclosed, analyzing CoT-style outputs can offer valuable insights into how the chatbot "sees" images.

### D. The role of multiple representations

Physics as a discipline relies on a range of different representations to capture and communicate disciplinary meaning. While algebraic expressions are often seen as the "language of physics," they are only a part of the rich representational repertoire that physics uses [75–77]. In physics education, this repertoire is sometimes expanded with representations that help learners develop conceptual understanding [78], even if they are later not used to the same degree by experts [79]. On the other hand, some non-algebraic representations, such as Feynman diagrams, have actually facilitated progress in physics research and remain integral to experts' work [5]. The topic of multiple representations is a well-developed research area in physics education [45]. Most commonly, the representations discussed in physics education research are either linguistic, mathematical, or graphical. However, some research has also looked at embodied representations (e.g., [6,7,80–82]). The topic of multiple representations in physics can be approached through different theoretical lenses, from the cognitive theory of multimedia learning [45] to a communication-centered approach of social semiotics [79], to name just two.
Regardless of the particular theoretical lens chosen, electromagnetism is an especially interesting topic for the study of multiple representations. Unlike classical mechanics, which often aligns closely with human everyday experiences, electricity and magnetism involve phenomena and concepts that are less directly accessible to the senses. Consequently, learning electricity and magnetism requires students to learn how to work with representations as tools for "making the invisible visible" [83], and enabling disciplinary reasoning on the topic. These representations include vector fields and circuit diagrams, alongside with those that students have already met in mechanics, such as force diagrams and graphs. Because electromagnetism often cannot be simplified to two-dimensional problems, there is also a need for three-dimensional representations, which can be fitted on a two-dimensional medium (i.e., paper or screen). To aid in visualizing and reasoning about three-dimensional relationships, physics has also developed some helpful heuristics. One of these is the right-hand rule, which allows for representing and reasoning about the direction of interrelated physics quantities, i.e., vector fields, velocities of charged particles, and forces experienced by charged particles.
Research shows that learning to use novel representations in electricity and magnetism is often difficult for the students [84,85]. The potential promise of LMMs is to aid students in developing the required representational competences by providing high quality, multimodal, personalized teaching [86].

## III. METHOD

### A. The Brief Electricity and Magnetism Assessment (BEMA)

To assess ChatGPT's ability to interpret multiple types of physics visual representations, we tested its performance on the Brief Electricity and Magnetism Assessment (BEMA)[2]. BEMA is an established and widely used research-based assessment of students' conceptual understanding of fundamental topics in electromagnetism, including electrostatics, magnetostatics, electric circuits, electric potential, and magnetic induction. Developed by Ding et al. [13], the test has been extensively used in physics education research(e.g., [88–90]). It consists of 31 multiple-choice items, 30 of which include visual representations such as images of electrical circuits, magnets and magnetic fields, point charges and electrical fields, force diagrams, current loops, sketches of experimental setups and graphs. In 14 items, the offered answer options themselves contain visual elements, such as arrows, graphs, or depictions of objects with charge distributions. This richness of visual representations makes BEMA well-suited for our purpose of investigating ChatGPT's ability to work with multiple physics representations. We take the results from a sample of 12214 students who took a university-level calculus-based electricity and magnetism course [14] as a reference point for comparison between the students' and chatbots' average performance on the test.

### B. Data collection

All responses were generated using OpenAI's online chat interface, with ChatGPT-4 tested in April 2024 and ChatGPT-4o in May 2024. We took high-resolution screenshots of all 31 BEMA items. Each screenshot captured the whole test item, comprising a visual representation of a situation, a question, and answer options. For 13 items, multiple survey items referred to the same image. For these, we combined all items referring to the same image into a single screenshot. In two cases, 3 items were captured in the same screenshot. As a result, the original 31 BEMA items were condensed into 22 screenshots. Figure 1 (left) shows an example of one such screenshot. The chatbot always answered all questions in a screenshot. Such an approach resembles the way in which the survey is presented to the students. The prompt given to the chatbot consisted solely of the uploaded screenshot, without any additional textual instructions.

---

[2] In order to protect the test integrity, we do not share the test. To facilitate the reading and interpretation of this paper, we suggest readers download the whole test from the Phyport.org website [87]. Some examples of images from the test are visible in the figures in this paper.

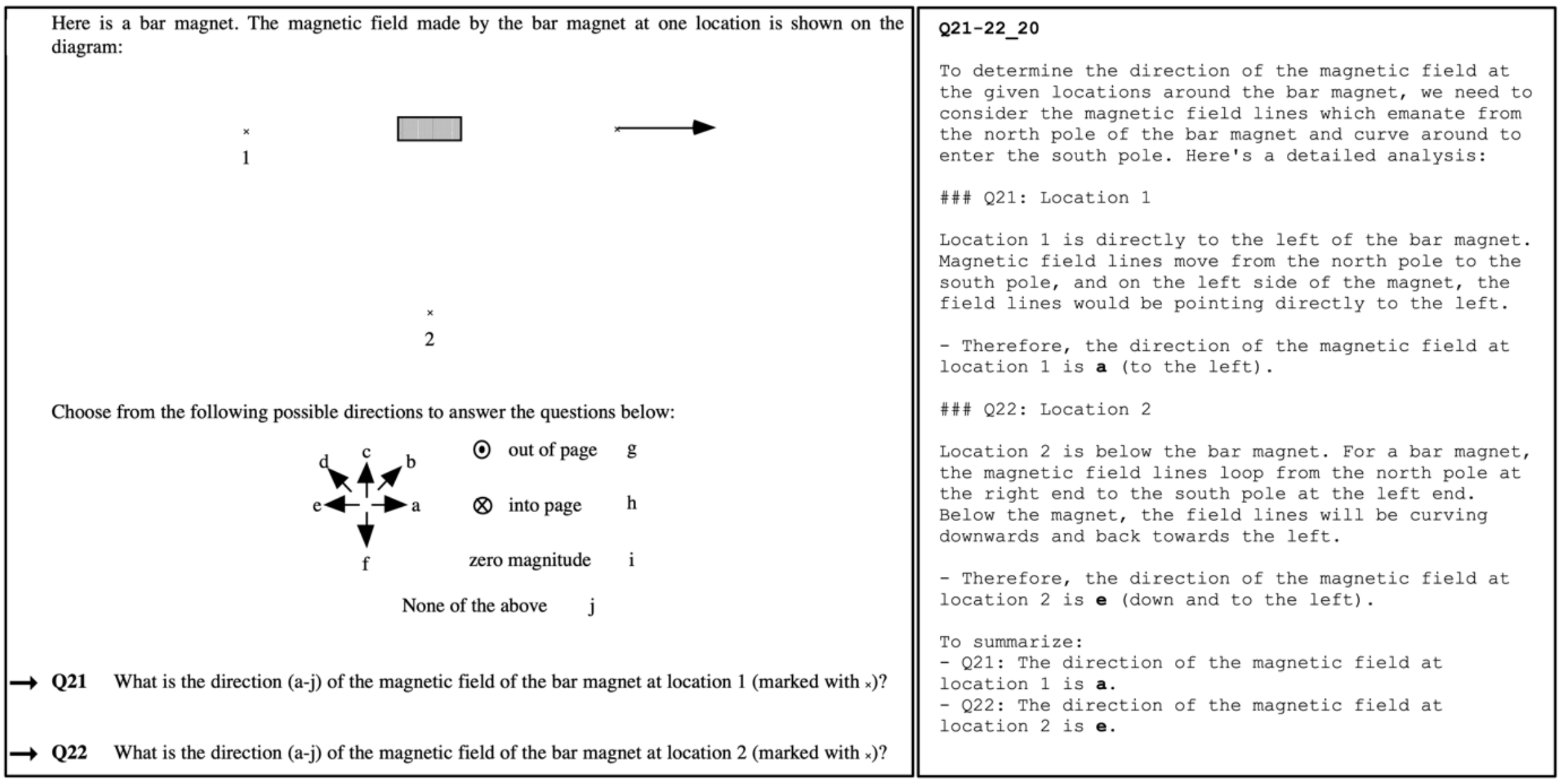

*Figure 1: Example of two related items in the same uploaded screenshot (left) and one of the responses to it, given by ChatGPT-4o (right).*

Each screenshot was submitted 60 times to ChatGPT-4. Because we prompted the chatbot in separate chat windows for each screenshot and disabled the “memory” function in the ChatGPT settings, we treated the performance on an item as an independent variable. Due to the probabilistic nature of chatbot outputs [25], responses were different every time, often arriving at different final answers on the same item. We determined the correctness of each response[3] and awarded it a score of either 0 or 1. This allowed us to treat the chatbot’s performance on each item as a Bernoulli variable, with an experimentally determined success rate corresponding to the chatbot’s performance score on the item (N of correct responses/N of iterations). The standard error (SE) of the performance score for each item was determined by taking the standard deviation of the item sample, divided by the square root of the number of iterations. The standard error for the different items was in the interval from 0.0 to 6.5%. The standard error of the mean (SEM) across all items on the survey was 0.8% of the available points on the test.

As a second step, we repeated the process with ChatGPT-4o. The only methodological difference is that we iterated each item 30 times instead of 60. We reduced the number of iterations as a compromise between time consuming data collection and the fact that the largest uncertainties for item perfoemance were only about three percentage points larger. For ChatGPT-4o the SE was thus in the range between 0.0% and 9.1% for the different items. The SEM across all items on the survey was 1.1% of the available points on the test.

[3] To simplify this process, we used the basic scoring key for individual items and did not use the advanced scoring approach for the test as a whole, which links together answers on certain items [87].

We have made the data available in an open online repository, as a resource for supporting potential future studies [71].

### C. Coding of answers

Like in our previous work [12,39,40], we coded the answers by noting the letter option stated by the chatbot, typically at the very end of its response (see the last four lines in Fig.1, right). In most cases, the chatbots provided a clear answer by explicitly stating a letter, making the process mostly straightforward. When a chatbot failed to select one of the provided choices explicitly or indicated that none were correct, we marked the response as not answered, "N," and considered it incorrect (5.5% for ChatGPT-4 and 3.2% for ChatGPT-4o).

However, we observed some discrepancies between the letter option stated as the answer and the content attributed to that letter option (see the response in Fig. 1, for example). To further investigate the impact of this inconsistency on the measured performance of the two models, we implemented a second coding approach. We looked at the content of the chatbots' output, and matched it to the corresponding answer option, regardless of the letter explicitly stated. For example, in Fig.1, using the second coding approach, the answer to question 21 was coded as "e," because it states "to the left", and the answer to question 22 was coded as "N" because it states "down and to the left", which did not match any of the offered answer options. For simplicity, we refer to the initial coding method as *coding by letter*, and to the second as *coding by meaning*.

### D. Qualitative analysis of difficulties

In Section V, we take a closer look at the difficulties exhibited by ChatGPT-4o, which is the better performing, more recent and computationally cheaper model. We decided to limit our attention only to the items with below-average performance, i.e., items on which less than 67% of the responses are correct, when coded by meaning. We analyzed 14 items, ranging in performance from 3% to 60% (items 4, 5, 8-10, 21-26, and 28-30). The chatbot's responses always contained step-by-step explanations, which allowed us to see the path by which the model arrived at the final answer and infer what difficulties it had (see section II.C for a discussion of the Chain-of-Thought approach).

In line with our previous research, we began by coding if a response contained any of the following two types of difficulties: difficulties with vision, and difficulties with physics reasoning [12]. However, in doing so, we observed a type of difficulty which did not fit into the two aforementioned two categories. Namely, we saw that the model often had issues with correctly solving the task, despite having correctly described the relevant features of the accompanying image (i.e., it did not exhibit difficulties in vision), and correctly stated physics principles/rules/procedures to be applied and applied them consistently (i.e., did not exhibit difficulties in reasoning). This type of difficulty can be described as unsuccessful spatial coordination of the described visual elements and stated physics laws/rules/procedures. For example, this happened when the model incorrectly applied a correctly stated the right-hand rule when determining the direction of the magnetic field around a wire with a correctly described orientation.

For each of the 420 analyzed responses (30 responses on each of the 14 items), we thus noted if it contained any of the following three types of difficulties:

1. Difficulties with vision. This category includes errors in visual interpretations of the representations accompanying the items (excluding the difficulties in interpreting the answer options field, accounted for in section IV). For example, the model had difficulties correctly describing the location of an element in the image.
2. Difficulties in providing correct physics laws or rules. For example, the model stated an incorrect right-hand rule.
3. Difficulties with spatial coordination and application of physics representations, laws, rules, and procedures. For example, the model incorrectly applied the principle superposition of electric fields in a given point, despite stating the principle correctly and describing the position of charges relative to the point correctly.

Two coders independently coded the 420 responses for the presence of each type of difficulty stated above. A response could contain more than one type of difficulty. Whenever there was any discrepancy in the coding of any of the three difficulty types, the response as a whole was marked as a disagreement. Disagreements were noted on 46 of the responses, making the intercoder agreement 89%, which is considered high and suggests a robustness of the coding system. The disagreements were resolved after a discussion and joint coding of the discrepant items.

In addition, we observed that the model had a lot of difficulties with three-dimensional problems that typically require the use of the right-hand rule. We additionally analyzed the responses to these tasks. For each response, we noted (1) if (a version of) the right-hand rule was explicitly stated, (2) if it was stated correctly (coupled the directions to each other in the correct way), (3) how it was formulated (e.g., what the thumb, fingers and palm stand for), and (4) if it was applied correctly in the accompanying task. This allowed us to glean further insight into the model's behavior on a subset of 7 items (23-26 and 28-30) that typically involve using the right-hand rule.

## IV. QUANTITATIVE ANALYSIS

In the first, quantitative part of our analysis, we answer RQ1 by examining the performance of ChatGPT-4 and 4o on BEMA in terms of their scores on individual items and across the entire test.

We do this by first comparing both chatbots' scores when we code the responses by the letter that the chatbot states at the end of every answer (*coding by letter*). We then examine how the score changes for each chatbot when we code their answers based on the verbal content of their answer (*coding by meaning*). The difference in each chatbot's performance gives us initial insights into their visual interpretation ability. We then compare the chatbots' performance when their responses are coded by meaning. This coding system better captures the chatbots' abilities to interpret and work with physics representations. Finally, we compare the performance of the better-performing and more recent vision-capable model, ChatGPT-4o, to that of a sample of physics students [14].

As outlined in Sec. III.D, we started coding both the answers from ChatGPT-4 and 4o by letter. According to the letter coding system, ChatGPT-4 achieves a performance of 50.2% on BEMA, while ChatGPT-4o reaches 59.8%. Notably, both chatbots performed better on BEMA than on the TUG-K, where the performance was 42.4% and 58.6%, respectively [40]. The higher performance on BEMA likely stems from the differences in the types of representations used on both tests.

### A. The difference in performance depending on the coding system

This mismatch between the two coding systems occurred in 26.8% of ChatGPT-4's responses and 10.8% of ChatGPT-4o's responses. The analysis suggests that most of the time this was caused by the chatbot misinterpreting the spatial arrangement of the answer option lists, mismatching a letter with the content belonging to another answer option due to their visual proximity. Usually, in items where directions were represented using arrows labeled with letters, the chatbot mismatched the verbal description of the direction with the corresponding letter (see Fig. 1). These types of answer option lists are present in 10 out of 31 BEMA items and their graphical design appears to have negatively impacted the performance. However, our analysis cannot exclude other possible explanations contributing to this mismatch, such as the biases of LLMs when selecting answers to multiple-choice questions [91]. Coding by meaning allows us to bypass the impact of this visual interpretation problem and measure the chatbots' ability to interpret other graphical elements in the items, namely the more canonical physics representations.

*1. The difference in performance of ChatGPT-4 when using the two coding systems*

When coding ChatGPT-4's responses by letter, the chatbot scored an average of 50.2% on the test. Passing from coding by letter to coding by meaning, its performance improved by 10.5 percentage points, indicating its ability to extract and process relevant information from graphical representations despite difficulties with answer option formats. Figure 2 presents the performance as captured by both coding systems. It also highlights the differences in scores on individual items: when the color of the range line is the same as "ChatGPT-4_meaning," it means that the performance has improved when switching to coding by meaning.

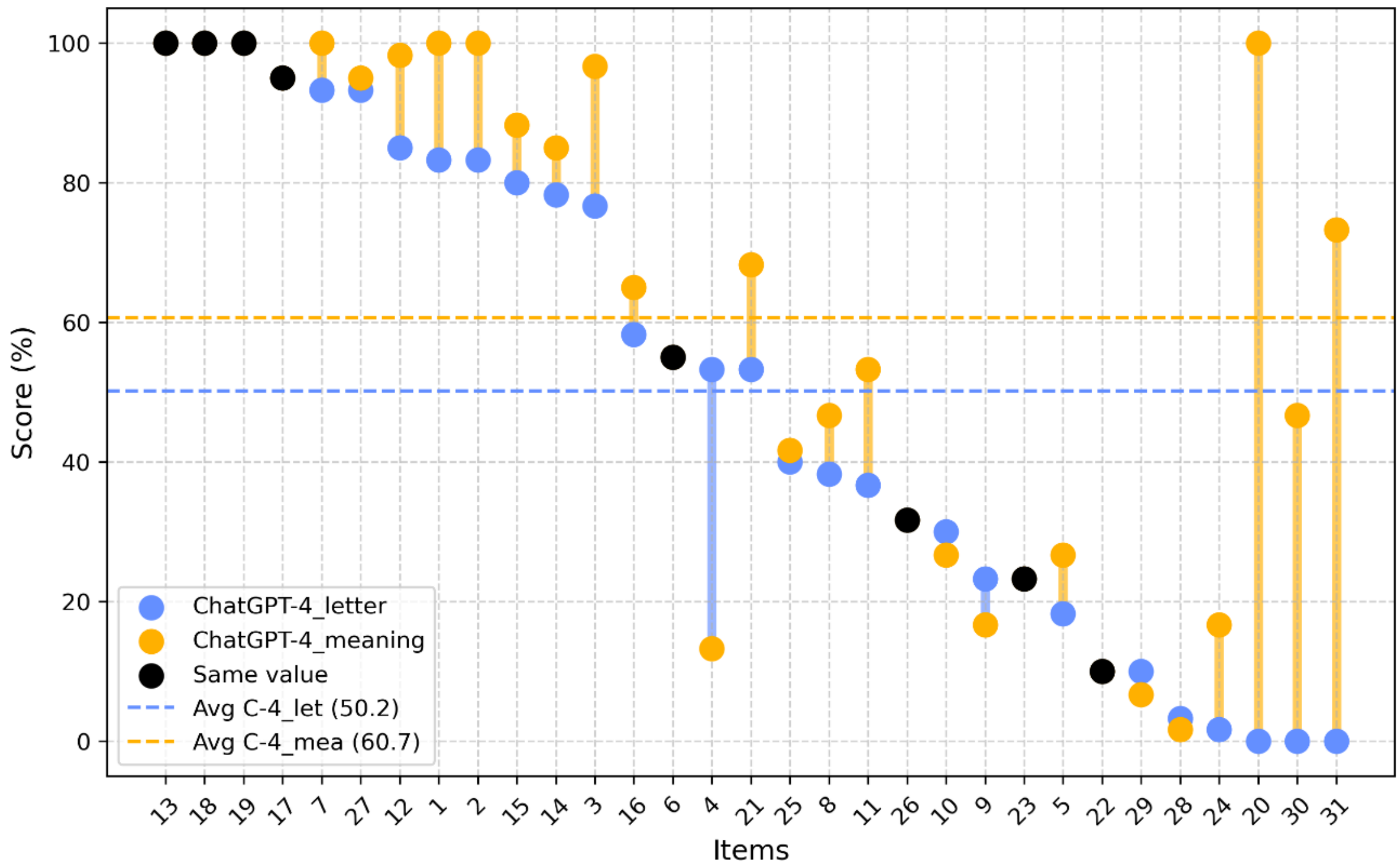

*Figure 2: Differences in the performance of ChatGPT-4, when the answers are coded by letter (orange) and meaning (violet). Test items are ranked from highest to lowest score according to the letter coding system.*

When coding by meaning, the performance is better for 18 out of 31 items. Item 20 performance improves the most in this shift, passing from 0% to 100% correctness. This item features an answer option layout very similar to that depicted in Fig. 1 (section III.B), with the issue consistently being the incorrect association of the selected letter with the intended meaning.

On the other hand, the performance is worse for 5 out of 31 items. For 4 of the items, the decrease in performance is 6 percentage points or less. Only item 4 stands out, showing a decline of 40 percentage points. However, a closer analysis shows that the correct letter was mostly selected by mistake, and coding by meaning reveals the chatbot's difficulty in correctly visually interpreting the image accompanying the item.

The performance on items 22, 24, 28, and 29 is under 20% in both coding systems. All of these items concern magnetism, with three requiring the use of the cross-product or right-hand rule. We will address this in more detail in the analysis of the responses from ChatGPT-4o, which turns out to have similar difficulties on these items (see section V).

*2. The difference in performance of ChatGPT-4o when using the two coding systems*

When coding ChatGPT-4o's responses by letter, the chatbot scored an average of 59.8% on the test. Switching to coding by meaning resulted in a 7.2 percentage point improvement, bringing the average performance to 67.0%. Figure 3 presents the scores for individual items and the overall difference in average performance across the two coding systems.

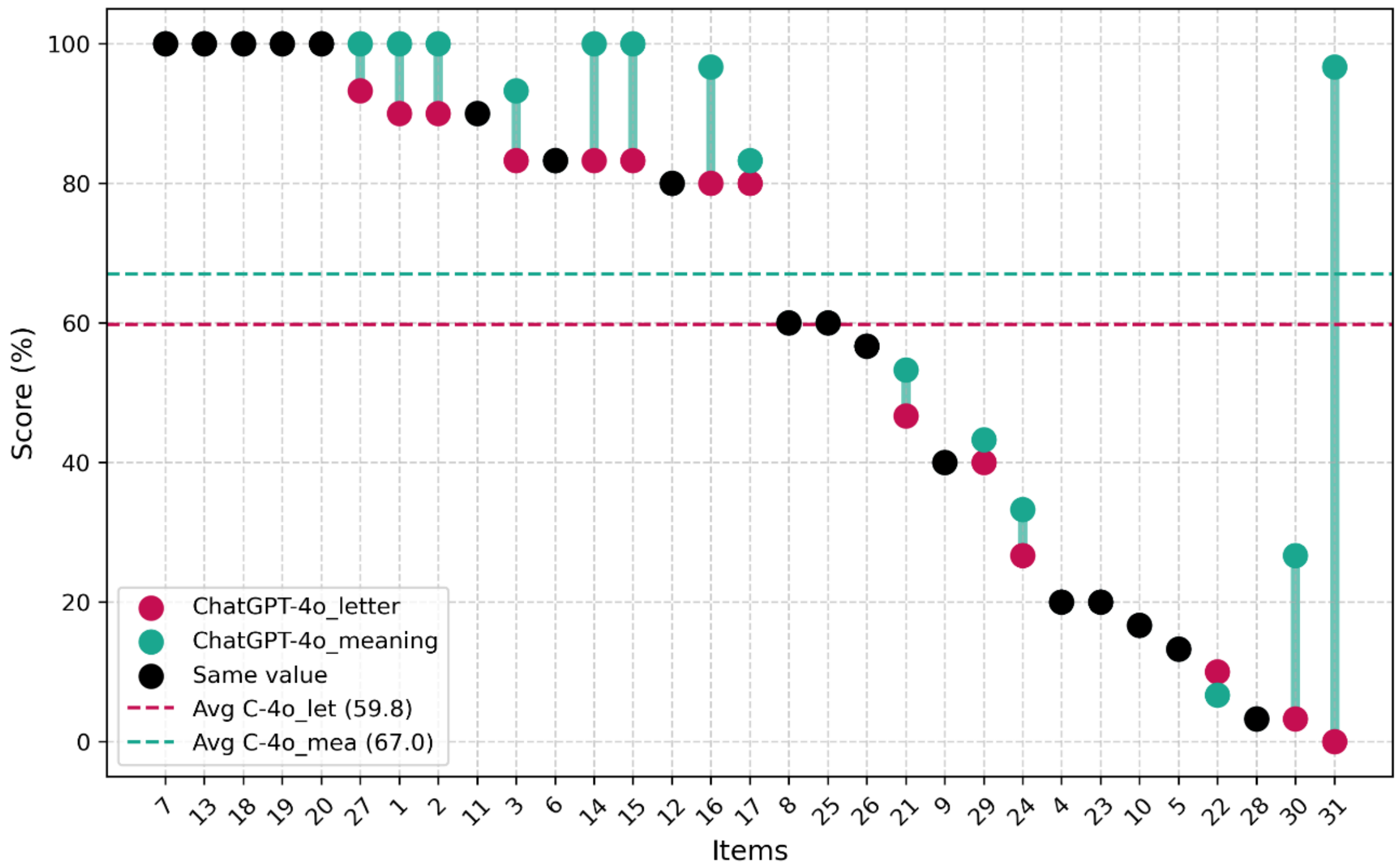

*Figure 3: Differences in the performance of ChatGPT-4o, when the answers are coded by letter (magenta) and meaning (green). Test items are ranked from highest to lowest score according to the letter coding system.*

Despite the improvement being much smaller than for ChatGPT-4, ChatGPT-4o's performance increases for 16 out of 31 items, when coded by meaning. The improvement is largest for item 31, accounting for 3.1 percentage points of the total 7.2-point improvement. Interestingly, this item includes six graphs as possible answer options, each depicting the time dependence of a voltmeter reading. Consistent with our previous research, ChatGPT-4o struggles with graph-based tasks, especially when they require selecting the right graph among several options [40]. In this case, the chatbot always correctly described how the graph should look like, but failed to pick out the graph that matched the description. In addition, 14 out of 31 items kept the same score, and only item 22 presents a small decline.

*3. Comparison between ChatGPT-4's and ChatGPT-4o's performance when coded by meaning*

From here on, we analyze the models in terms of their performance when the answers were coded by meaning. The comparison between ChatGPT-4 and 4o, shown in Fig. 4, disregards the multiple-choice answer selection step and takes into account only the answer arrived at through reasoning. This bypasses potential difficulties stemming from the format of the answer options list, or potential biases related to the indexing of multiple-choice answers [91]. The performance coded by meaning thus more directly tracks the chatbots' abilities to work with physics representations embedded in the test items.

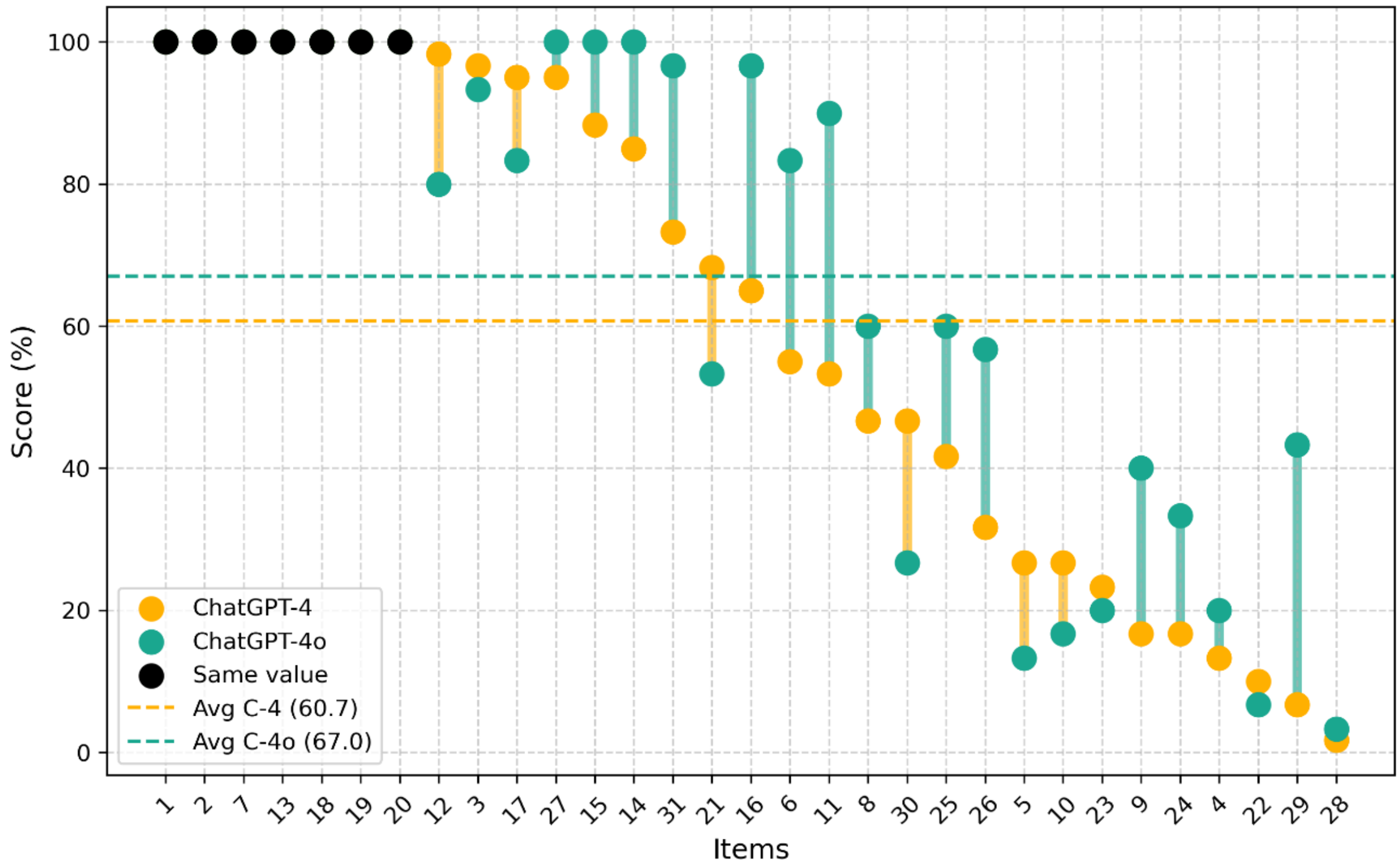

*Figure 4: Difference in test performance of ChatGPT-4 (orange) and ChatGPT-4o (green), when the answers are coded by meaning. Test items are ranked from highest to lowest score according to ChatGPT-4.*

When coding by meaning, ChatGPT-4o outperforms ChatGPT-4 on nearly half of the items, while ChatGPT-4 performs significantly better on six items. We could not identify any meaningful pattern that would explain the differences. Although the difference in both models' performance on the test as a whole is only 6.3 percentage points, it signifies an advantage in ChatGPT-4o's ability to handle tasks involving the interpretation of graphical physics representations.

### B. Comparison between ChatGPT-4o's and students' performance

Previous research on ChatGPT-4o's performance on visual tasks indicates that it outperforms both its predecessor ChatGPT-4 [40] and students at the high school level [21,39]. On BEMA, its performance, regardless of the coding system used, exceeds the average score for students (53.4%) as reported by Wheatley et al. [14] for a sample of 12214 students enrolled in an introductory calculus-based electricity and magnetism course. Figure 5 shows the comparison between its performance, coded by meaning, and that of the student sample.

A notable difference between ChatGPT-4o and the student dataset is the distribution of item scores. Student scores were relatively evenly spread out between 16% and 83%, with two items at 80% or above, and 3 items at 20% or below. In contrast, the chatbot's distribution was more tail-heavy, with 14 out of 31 items scoring 80% or higher (including 10 items at 100%) and 6 scoring 20% or less. This pattern, similar to that previously observed in other work [39,40], highlights the chatbot's tendency to perform either very well or poorly on specific items. We can also see that on items 4, 5, 8, 10, 21-24, 28, and 30, students on average outperform ChatGPT-4o. All of these items saw below-average performance from ChatGPT-4o and were included in the qualitative analysis presented in the following section. ChatGPT-4o's performance in relation to students' performance is further discussed in section VI.B. Among the items where ChatGPT-4o outperforms students, those that exhibit the largest gap (> 50 percentage points) are items 12, 14, 16, 17, 27, and 31. All these items require formal and mathematical reasoning, which are known to be challenging for students. Indeed, the students' performance on these items is poor (28% on average). On the other hand, ChatGPT-4o is relatively successful on these tasks (93% on average).

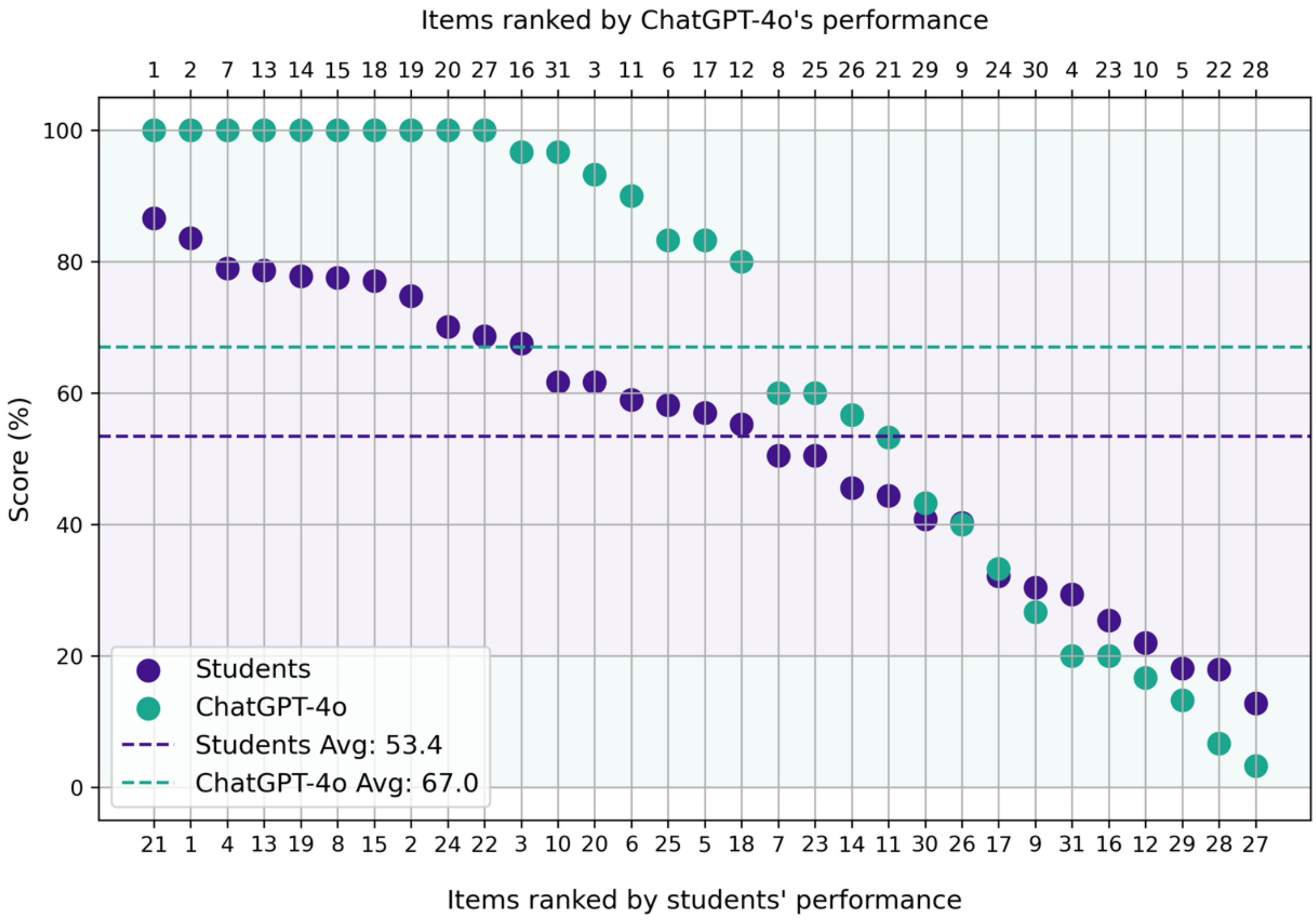

*Figure 5: Test performance of ChatGPT-4o (green) and a sample of students [14](purple). Test items are ranked from highest to lowest score for students (bottom) and ChatGPT-4o (top).*

## V. QUALITATIVE ANALYSIS

In the previous section, we presented the findings of the quantitative analysis of the performance of ChatGPT-4 and 4o on BEMA. This section focuses on answering the RQ2 through a qualitative analysis of a subset of ChatGPT-4o's responses. We limit our focus to ChatGPT-4o, which is the more recent and better-performing of the two chatbots. Additionally, it is faster and cheaper to run, thus likely to be used more broadly in educational applications.

In all its responses ChatGPT-4o provided a step-by-step explanation of the steps taken in solving each item. As we have explained in section II.C, this feature, also referred to as the Chain-of-Thought approach, is not only useful for users who want to understand how the answer can be arrived at, but is also a strategy employed to improve the model's performance. Furthermore, it also provides us with a window into its processing of the tasks we submit to it.

To better understand its difficulties, we focus our analysis on items on which the chatbot showed below-average performance when coded by meaning. For simplicity, we draw the line for poor performance at average performance (67%), giving us 14 items ranging in performance from 3% to 60%. We analyzed each of the 30 responses on these 14 items (420 in total), noting down observed difficulties.

We categorized the observed difficulties into 3 categories, described below. We provide an example of each of these difficulties, as observed in our data, and a quantitative summary of their prevalence. Additionally, in section V.D, we further narrow our focus on a subset of seven tasks requiring three-dimensional reasoning about moving charges/currents and associated magnetic and electrical fields, which typically involve the use of the right-hand rule.

## A. Difficulties with vision

Despite ChatGPT-4o showing an improvement over ChatGPT-4 in visual interpretation, a number of issues remain. Of the 420 responses, 136 (32%) contain errors in visual interpretation. Most commonly, these are difficulties in correctly describing the location of points of interest in the image (43% of responses on items 4, 5, 21, 22, 28, and 29). See Fig. 6 for an example. Other difficulties include incorrectly describing circuits (83% of responses on item 10, see Fig. 7 for an example) and incorrectly identifying the direction of current in a loop shown in perspective view (when the plane of the loop is not the same as the plane of the page) (70% of responses on item 24).

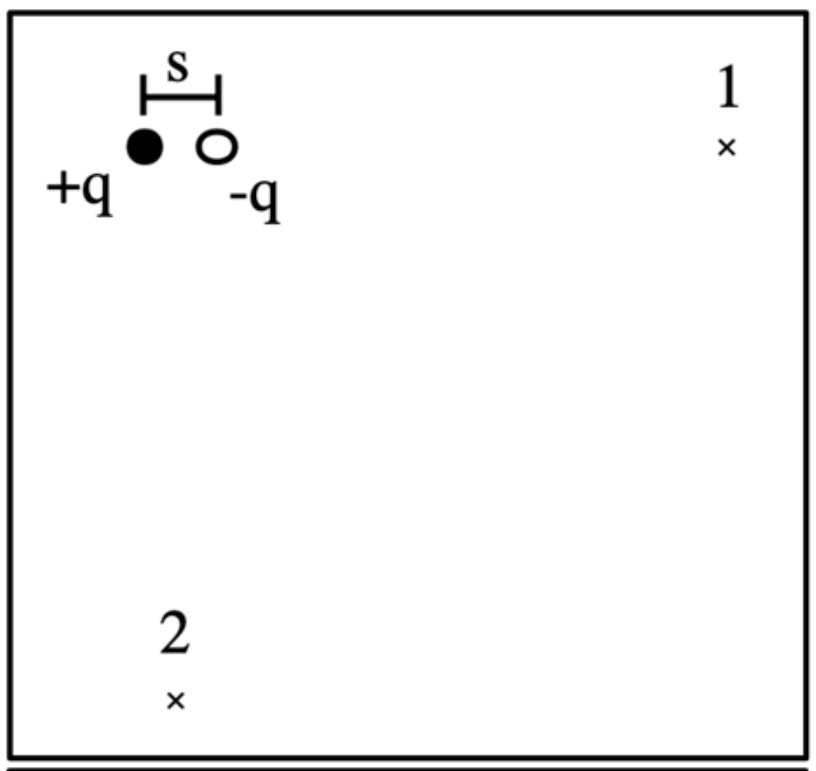

[…]### Location 1

Location 1 is positioned directly to the right of the +q charge and directly above the -q charge.[…]

*Figure 6: Image shared by items 4 and 5 (above) and the relevant part of ChatGPT-4o's description of the image (below). The description is incorrect.*

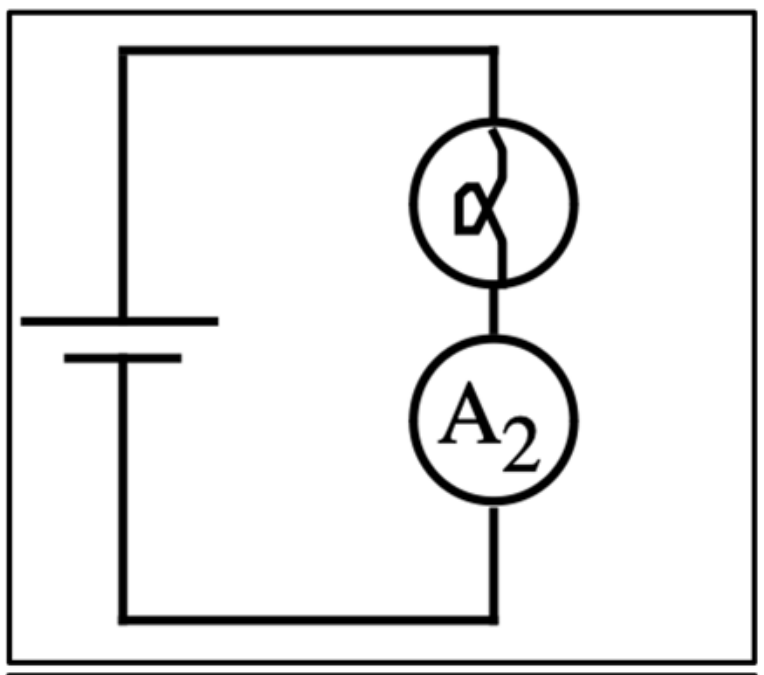

[…]2. **Circuit with $A_2$**:

- This circuit consists of only a battery and an ammeter $A_2$.

- There is no resistance other than the internal resistance of the battery and the ammeter itself.[…]

*Figure 7: A cropped part of the image in item 10 (above) and the relevant part of ChatGPT-4o's description of the image (below). The description is incorrect.*

## B. Difficulties with providing correct physics laws or rules

This category includes errors such as incorrectly stating physics laws and principles, or incorrectly deriving their consequences. This type of difficulty was the least common of the three categories and was present in 14% of the analyzed responses. For example, it was identified as the main culprit of wrong responses on item 9 (see Fig. 8). The relatively low number of errors in this category is not particularly surprising since language models are typically well-equipped for declarative knowledge of physics laws and principles. This corroborates our previous findings [39,40]. Still, the 14% prevalence of this type of difficulty is not insignificant, indicating that even in areas where ChatGPT-4o is typically expected to excel, inaccuracies can occur.

The magnitude of the drift velocity of the sodium ions is $v_{Na}$, and the magnitude of the drift velocity of the chloride ions is $v_{Cl}$. Assume that $v_{Na} > v_{Cl}$. (+e is the charge of a proton.)

→ **Q9** What is the magnitude of the ammeter reading?

(a) $enAv_{Na} - enAv_{Cl}$
(b) $enAv_{Na} + enAv_{Cl}$
(c) $enAv_{Na}$
(d) $enAv_{Cl}$
(e) zero

**Since the chloride ions are negatively charged and move in the opposite direction to the sodium ions, their contribution to the total current will be in the opposite direction. Therefore, the total current *I* is the difference between the current due to sodium ions and the current due to chloride ions.**

**The total current *I* can be written as:**

$I = I_{Na} - I_{Cl} = neAv_{Na} - neAv_{Cl}$

**Simplifying, we get:**

$I = neA(v_{Na} - v_{Cl})$

****Answer: (a) $enAv_{Na} - enAv_{Cl}$ ****

*Figure 8: Question from item 9 (above) and the relevant part of ChatGPT-4o's response (below). The reasoning is incorrect.*

## C. Difficulties with spatial coordination and application of physics representations, laws, rules, and procedures

ChatGPT-4o frequently struggles with correctly applying physics principles (for example, the superposition principle and the right-hand rule) in the context of the concrete spatial configuration in a task. For example, even when it correctly stated the principles for finding the total field of an electrical or magnetic dipole, and correctly located the points of interest, it often failed to execute the procedure of obtaining the total field in a given point (see Fig. 9 for an example). This appears to be due to its difficulty in sustaining and applying a coherent spatial model of the situation, often confusing up and down, left and right, and into and out of the page (see Fig. 10 for an example). This type of difficulty was present in answers to all of the survey items, except items 8, 9, and 10, and in 60% of responses on items 4, 5, 21, 22, 23, 24, 25, 26, 28, 29, and 30, making it the most prevalent type of difficulty in our data.

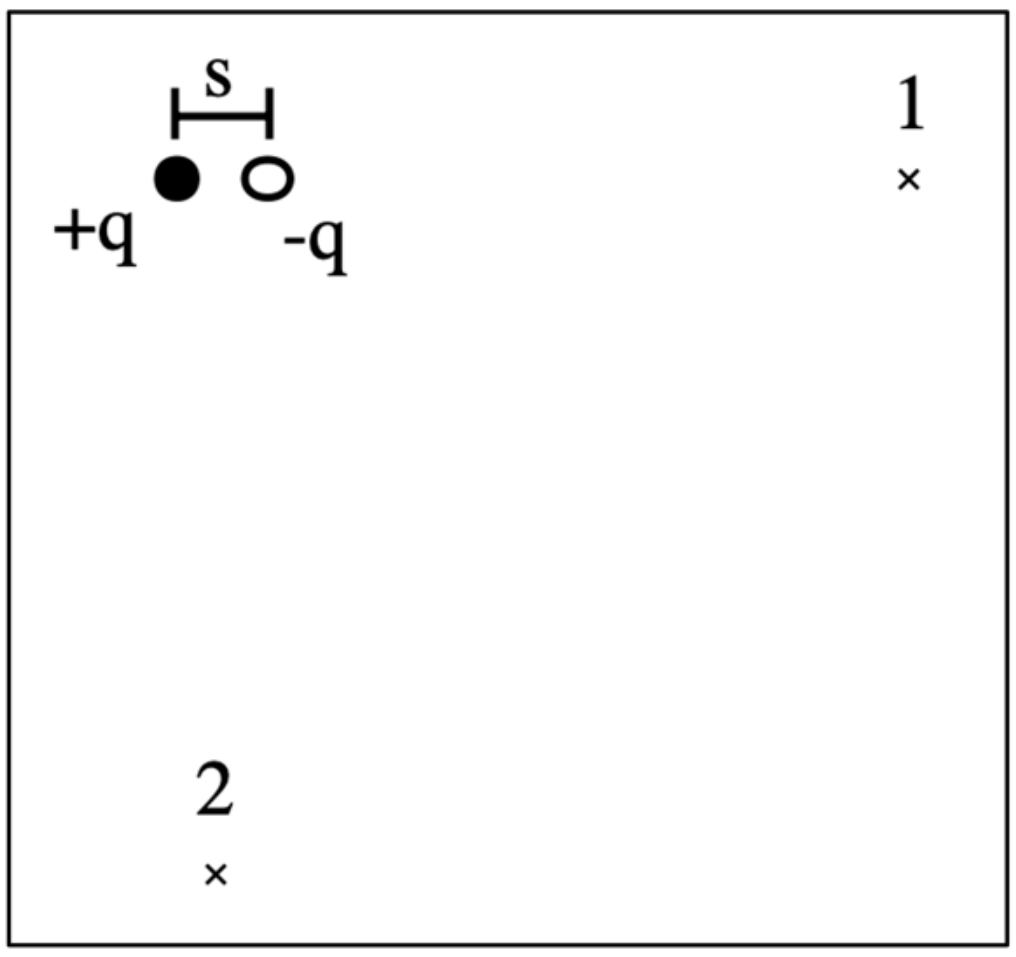

[…]### Configuration:

We have two charges:

+q on the left

-q on the right

Points to analyze:

- **Location 1** is directly to the right of -q

- **Location 2** is directly below the midpoint between +q and -q […]

[…]### Electric Field at Location 2:

**From +q**: The field points away from +q downwards and rightwards.

**From -q **: The field points towards -q downwards and leftwards.

At location 2, both fields have a vertical component downwards and horizontal components cancel out due to symmetry:

Thus, the resultant field is directly downwards (f).[…]

*Figure 9: Shared image for items 4 and 5 (left) and the relevant parts of ChatGPT-4o's response to item 5, which asks about the direction of the electric field at point 2. Despite the correct description of the image and the superposition principle, the chatbot incorrectly determined the direction of the field contribution of the negative charge, which led to an incorrect final answer.*

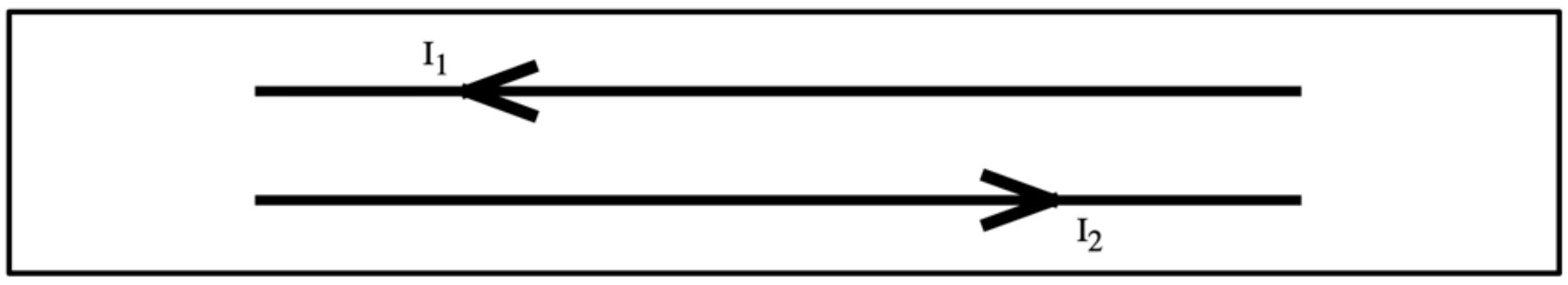

[…]To determine the direction of the magnetic force that wire 1 exerts on wire 2, we can use the right-hand rule for magnetic forces between parallel currents.

1. **Right-Hand Rule for Magnetic Field**: The direction of the magnetic field created by a current-carrying wire can be found using the right-hand rule. For a wire with current flowing to the left (wire 1), point the thumb of your right hand in the direction of the current, and curl your fingers. Your fingers curl around the wire in the direction of the magnetic field.

2. **Magnetic Field Around Wire 1**: For wire 1 (current $I_1$ flowing to the left), the magnetic field around wire 1 at the location of wire 2 (below wire 1) will be into the page (⊗).[…]

*Figure 10: Image from item 25 (above) and the relevant part of ChatGPT-4o's response (below). The description of the image is correct, and the right-hand rule is stated correctly, but the conclusion is incorrect, implying that the chatbot has difficulties with coordinating its vision and reasoning abilities when applying them to concrete situations.*

## D. Right-hand rule

Tasks on which the right-hand rule (RHR) can be used turned out to be especially interesting for our qualitative analysis (items 23-26 and 28-30). The average performance across these tasks was 35%, indicating that they are difficult for the model to solve. Some of these items contained subtasks (steps to solving the problem), each requiring the use of some form of the RHR (items 25, 28, and 29). Throughout the tasks and subtasks where the RHR could be used, it was explicitly invoked and explained in 53% of cases. When the RHR was not explicitly stated by the chatbot, an alternative rule was typically given, such as the cross-product formula connecting the directions of velocity (or current), the magnetic field, and force, or a statement about the magnetic field's direction in a clockwise or counterclockwise current loop.

Of the responses where some form of the RHR was invoked, 9% contained an incorrectly stated RHR, and 41% contained an incorrect implementation of the stated rule (see Fig. 10 for an example).

The chatbot articulated multiple versions of the RHR depending on the task type. Below, we outline these versions, their prevalence, and ChatGPT-4o's success rates.

*1. RHR for force on charged particle/current carrying wire*

This version of the RHR was explicitly stated in 79% of the responses on the four items where it could be used. When explicitly stated, it was correctly stated 95% of the time. However, the chatbot failed to correctly apply it in 70% of these cases, suggesting that it struggles with correctly applying this three-dimensional representation. Notably, humans typically do not perform this task in the "mind's eye," but rather *off-load* it to the body [92,93], using the right hand as an embodied representation for coordinating the graphical representation of the problem and the physics laws in three-dimensional space [84]. Not having a body [94], the chatbot does not have the means for such *off-loading*.
A further intriguing observation is that in some cases, the model of the hand inferred from the chatbot's output was either anatomically impossible, or it referred to the left hand, as illustrated in the following quote from one of its responses to item 23:

> "Point your fingers to the right (direction of $v$). To get the thumb to point upward (force direction for a positive charge), your fingers would need to curl out of the page."

We also observed that different variants of the RHR for force on a charged particle/current carrying wire were described by the chatbot on repeated iterations of one and the same task. For example, the fingers, thumb and palm did not always stand for the same physics quantity. While in most cases the relative directions of the three quantities were correctly captured by the stated rule, this inconsistency is something that educators should take into account when planning on using ChatGPT-4o in educational contexts, or directing the students to interact with it.

*2. RHR for a current loop*

This version of the RHR relates the direction of the current in a current loop with the resulting magnetic field. It was explicitly stated by the chatbot in 30% of responses on the two items where it could be used. Of these, it was correctly stated 94% of the time. However, more often in the same tasks, the direction of $B$ was expressed without referring to the hand at all. Instead, the direction of the magnetic field was simply stated to be in or out of the page, depending on whether the current was circulating clockwise or counterclockwise (55% of responses on the two items). ChatGPT-4o was mostly successful in applying this rule on tasks where the loop was aligned with the plane of the page, but struggled when it was oriented perpendicularly to the page or shown in perspective view. This is somewhat similar to the difficulties experienced by students [84], reinforcing the idea that ChatGPT-4o has difficulties applying this three-dimensional representation in concrete tasks.

*3. RHR for magnetic field around a current carrying wire*

This version of the RHR was explicitly stated in 63% of responses on one item where it could be used. When stated explicitly, it was correct 95% of the time. Similarly to the other versions of the rule, it was applied unsuccessfully in 68% of cases where it was stated correctly (see Fig. 10 for an example). This finding further supports our finding that the ChatGPT-4o has difficulties working with three-dimensional representations and models of physics phenomena.

## VI. DISCUSSION

We divide our discussion into three parts. First, we comment on the technical aspects of our findings related to the performance of both tested LMMs and briefly reflect on the possible reasons for the difficulties exhibited by ChatGPT-4o. Second, we look at how our findings can inform the use of the chatbot in the three possible educational roles proposed in our background section. Third, we note some limitations of our study and propose possible future research directions.

### A. ChatGPT's performance on BEMA

We have found that ChatGPT-4o outperforms its predecessor, ChatGPT-4, exhibiting better performance on BEMA, regardless of the answer coding approach (by letter or by meaning). Additionally, ChatGPT-4o exhibits less discrepancy between the two coding approaches. One possible explanation for this is due to its improved ability to interpret the visually complex answer options list. Another possibility, suggested by Zheng et al. [91], is that it is less biased toward specific answer indexes.

Overall, our quantitative findings support previous research that positions ChatGPT-4o among the best LMMs for physics-related visual tasks [39,40]. Because it offers a good balance of reasoning and visual capabilities at a relatively low cost compared to newer and more advanced models, it will likely remain in broad use for some time to come, especially in third-party educational applications that access it via API. In fact, it is uncertain if OpenAI's newer LMMs, such as o1 [41], have improved their core vision abilities over ChatGPT-4o. A quick test of the most problematic BEMA items, performed since this paper's submission, suggests that o1 produces more elaborate reasoning, but still exhibits very similar visual errors. While these are only anecdotal findings, systematic research is needed to compare the vision capabilities of o1 against 4o.

Our qualitative analysis identified three qualitative categories of failure, which allow us to infer some difficulty patterns across our dataset. However, classifying tasks in more detail based on their difficulty in relation to specific representational forms (sketches, diagrams, graphs, etc.) or involved procedures (determining total resistance, finding the direction of a field by the superposition principle, etc.) is much more challenging. One reason is that our data consists of only 31 survey items, with many non-systematically varied parameters. For example, to infer more precisely what visual features influence the performance of an LMM, a study would need to systematically vary an enormous number of graphical parameters, such as line thickness, spacing, positioning, and orientation of elements, font size, and many more, not all of which are necessarily known in advance to be relevant. Furthermore, a human-centric conceptualization of different representational features may fail to capture the factors that most strongly impact LMMs' visual processing. In previous studies [39,40], we examined different chatbots' performance on tasks involving kinematics graphs, which represent an area of much narrower representational diversity than electricity and magnetism. Still, we could not infer many regularities based on content or procedural classification of different tasks and saw that tasks that appear very similar to the human eye could exhibit very different difficulties for chatbots. We hypothesize that this is mainly due to the mechanisms behind LMMs' visual processing of the tasks, which are different from those in humans. We suspect that in the process of embedding images for further processing, some information about the details in the image can be lost [95]. The LMM, not being aware of this, fills in the missing information according to statistical regularities in its training data. This can result in uncanny descriptions of images, which are blatantly wrong. This can, to some extent, be seen by a simple examination of the type of errors exhibited by ChatGPT-4o in our data. We further address this matter in the next section, where we discuss how our findings can inform the use of ChatGPT-4o in different educational roles [46].

One particularly challenging representation for the chatbot was the right-hand rule (RHR). The RHR is an embodied heuristic that allows humans to off-load to the body the spatial cognitive task of cross-multiplying vectors [84]. However, the RHR does not seem to play the same role for ChatGPT-4o. This is not surprising, given that the processing of information by the chatbot happens on the level of language, and does not involve manipulating an actual three-dimensional model of a hand. For humans, it is mostly trivial to tell where the fingers on one's hand are pointing once the hand is physically positioned in space. However, ChatGPT-4o seemed to have a lot of difficulties inferring this information from its verbal descriptions of the RHR applied in concrete situations. This finding is relevant not only from a physics instructional perspective – one cannot rely on it in tasks involving the RHR – but also reveals an important underlying limitation of ChatGPT-4o to reason spatially.

To conclude the discussion of the technical aspects of our findings, we consider possible ways in which the performance of LLMs could be improved. Training new models on large and diverse databases of labeled visual physics representations will likely be an important component in improving their vision abilities, as well as fine-tuning existing ones. More specialized multimodal models, trained primarily on scientific visualizations and representations, may also prove advantageous. However, this is not something that can typically be done by individual educators, due to the high demands on computational capacities and amounts of data that such training would entail. Another possibility that has not yet been systematically explored in the context of physics, is examining whether specialized prompting approaches aimed at "guiding the LMM's vision" to relevant aspects of a given representation can influence the way images get processed. As such attempts to improve LMMs' performance take place, it is important to keep assessing them. Our study provides one possible methodological approach for such assessment, which is flexible enough to be used across different domains of physics and associated representational forms.

### B. Implications for instruction

Here, we relate our findings to Taylor's three roles framework [46] and discuss how they may inform the use of ChatGPT-4o in the different educational roles.

#### *1. ChatGPT-4o as a physics tutor*

Since the first release of ChatGPT, OpenAI's LLM- and LMM-based chatbots have seen significant improvements, particularly in reasoning abilities [17,40]. For educational uses in high-school and introductory-level physics, these improvements make ChatGPT-4o a reasonably capable tutor for many topics, as prior studies on its physics performance suggest. On the other hand, researchers also mostly agree that its multimodal performance is still unreliable (see section II.A). Our findings strengthen this and suggest that ChatGPT-4o's performance is not yet at the level that should be expected of a physics teacher.

Based on the difficulties described in this paper, we see significant potential for confusion if students rely on ChatGPT-4o as the primary or sole tutor to help them work with visual representations in electricity and magnetism. Learning to interpret novel representations is difficult enough, even without receiving conflicting information from the AI tutor. Such inconsistencies can be the result of both vision and coordination difficulties that we have identified in this study. A key example that we found is the chatbot's inability to consistently correctly apply the RHR. First of all, it frequently presented several different variants of the RHR, even for the same task. At times, the fingers stood for the direction of particle motion, the palm for the direction of the magnetic field, and the thumb for the direction of the force on a positively charged particle. Other times, the thumb stood for the direction of motion, the fingers for the direction of the magnetic field, and the palm for the direction of the force. While the different variants of the RHR were mostly correctly stated, different students could, in the best-case scenario, walk away from interacting with ChatGPT-4o with different understandings of the RHR. More likely, they would become confused about how to use the RHR. Teachers should be aware of the variety of ways in which ChatGPT-4o expresses the RHR and be prepared to assist their students in making sense of its output. We suggest students should be careful when letting ChatGPT-4o lend them a hand with tasks involving the use of the right-hand rule.

All in all, we would advise against having students uncritically use the chatbot as their main or even sole tutor on the topic of electricity and magnetism, especially when visual and spatial representations are explicitly involved. Teachers can use the findings of this paper to inform beginner learners about the limitations of this technology and encourage them to always cross-check information with verified sources. For more advanced learners, who can recognize and mitigate the chatbot's shortcomings, ChatGPT-4o can still represent a valuable tutor, mostly because of the predominantly productive strategies it suggests for solving physics problems at this level.

### 2. *ChatGPT-4o as a model of a student*

As we lack corresponding student reasoning data, we are unable to systematically compare students' and chatbots' approaches to solving different tasks on BEMA. However, we can see that many errors made by the chatbot would be quite atypical for human students, based on our experience as physics instructors.

One striking example is its difficulty in identifying spatial relationships in two-dimensional diagrams. While seemingly trivial, the chatbot frequently failed to correctly locate (i.e., left, right, above, below) a marked point in relation to another. Such errors are uncharacteristic of students. A closer examination of a group of items on which students outperformed ChatGPT-4o reveals that vision and spatial coordination difficulties are key contributors to its weaker performance. Possible reasons for this are discussed in the previous subsection. At the same time, the chatbot's responses typically contain detailed and well-formulated strategies for approaching the problem at hand. This uncanny combination of nearly perfect strategy descriptions and seemingly trivial vision and/or coordination errors can be seen as a telltale sign that a response was probably AI-generated. These findings reveal potential limitations in ChatGPT-4o's utility as a model of a student, e.g., for generating synthetic data of student responses to tasks involving visual representations. It also limits its potential to convincingly simulate struggling students for the purpose of teacher training through Socratic dialogue [23]. It is worth noting that directed prompting approaches have previously shown some promise in this regard [25]. For example, an LLM could be prompted to respond like a student having a particular difficulty [19]. Future studies still need to explore how different prompting approaches might impact LMMs' behavior in relation to visual inputs, in order to see if it can be made to behave more like a human student.

On the other hand, the limited visual abilities may also be leveraged in designing assessments that are inherently difficult for chatbots to solve. Tasks that employ multiple representations can be a strategy to reduce the usefulness of chatbots when their use is not desired, such as during exams. Again, some characteristic errors made by ChatGPT can also be a telltale sign that it was used in answering a question.

### 3. *ChatGPT-4o as an educational tool*

The shortcomings of ChatGPT-4o suggest that, despite improvements over ChatGPT-4, its usability as an educational tool in vision-heavy tasks is still limited. In combination with previous findings on the models' difficulties with interpreting kinematics graphs [12,40], we can say that the unreliability of ChatGPT-4 and 4o's performance is still too large for them to be useful as trustworthy educational tools at this point in time. This also extends to potential uses as accessibility tools for users with low vision. However, the shortcomings can still be leveraged by teachers. They can use ChatGPT to generate material for fostering critical discussion in the classroom, educating students on the importance of critical reading and evaluating the reliability of AI-generated information [69].

## C. Limitations and future work

The improvement in ChatGPT-4o over ChatGPT-4 illustrates the trend of ongoing advancement of the technology's abilities. Against this background, our work should be seen as a momentary snapshot and a historical reference point. In this study, we have focused our attention only on OpenAI's LMM-based chatbots, which were previously shown to outperform competitors on multimodal physics tasks [40]. However, we did not test other chatbots on BEMA and cannot compare their performance to that of ChatGPT. This work remains to be done.

Our findings are based on a specific type of assessment – a multiple-choice test on a particular topic. While the range of representations in electricity and magnetism is broad, the findings may not directly apply to other areas of physics and formats of assessment. Nevertheless, our approach to assessment is general and flexible enough to serve as a possible methodological stepping stone for future explorations.

However, as we have already discussed above, determining more specifically what features of a representation makes a task difficult for an LMM to solve would require a more systematic variation of many potentially relevant parameters over a large number of different input images.

In this study, we did not use any specialized prompt-engineering techniques to try to enhance the chatbots' performance. Directed prompting approaches (e.g., instructing the chatbot to describe the images in great detail before solving the accompanying task [12]) may produce better outcomes. The area of prompt engineering in multimodal prompts has not yet received much attention and remains an exciting possible avenue for future work.

Another promising avenue for future research is exploring how disembodied AI systems perform on tasks where humans rely heavily on embodied cognition [6,96]. Our findings regarding the difficulties that ChatGPT-4o experienced when applying the right-hand rule can serve as a possible point of departure for such exploration.

## VII. CONCLUSION

Our quantitative analysis of OpenAI's two large multimodal models on BEMA showed that when coded by the selected letter option, ChatGPT-4's (50.2%) and ChatGPT4o's (59.8%) average performance across all items is similar to that of the average of a large sample of college students taking a calculus-based introductory course in electricity and magnetism (53.4%). However, we have also seen that both chatbots' performance is better (60.7% and 67.0%, respectively) when coding their response by the word content of the answer expressed in their reasoning. A qualitative interpretation of the answers suggests that one possible explanation of this mismatch is the difficulty that both versions of ChatGPT exhibit in correctly visually interpreting the area containing different answer options. Other possible explanations include inherent bias in the model, causing it to systematically prefer certain answer options. Our qualitative analysis of ChatGPT-4o's difficulties on BEMA reveals that in addition to inaccurate visual interpretation, a major difficulty was spatial coordination of different representations and physics principles. We found ChatGPT-4o's use of the right-hand rule to be especially problematic in this respect. We suggest that educators should be cautious when using ChatGPT-4o for educational purposes, especially in topics that require extensive use of graphical representations. This includes using it as a tutoring system, a model of a student, or an educational tool, such as an accessibility assistant for students who are blind or have low vision. On the other hand, knowledge of the identified shortcomings can be useful for the design of chatbot-resistant assessments.